\pdfoutput=1 
\documentclass[prl,twocolumn,preprintnumbers,superscriptaddress,showpacs]{revtex4-1}

\usepackage{graphicx}
\usepackage{bm}
\usepackage{amsmath}
\usepackage{amssymb}
\usepackage{slashed}
\usepackage[colorlinks,pdfstartview=FitH]{hyperref}
\hypersetup{linkcolor=blue,citecolor=blue,filecolor=black,urlcolor=blue}
\newcommand\sect[1]{\emph{#1.}---}

\newcommand{\no}{\nonumber\\}

\def \be {\begin{equation} }
\def \ee {\end{equation}}

\def \bem {\begin{multline}}
\def \eem {\end{multline}}

\def \bes {\begin{subequations} }
\def \ees {\end{subequations}}

\def \pd {\partial}

\def \o {\omega}
\def \O {\Omega}

\def \r {\rho}
\def \l {\lambda}
\def \m {\mu}

\def \s {\sigma}

\def \th {\theta}

\def \D {\Delta}
\def \G {\Gamma}

\def \<{\langle}
\def \>{\rangle}
\def \+{\dagger}
\def \({\left(}
\def \){\right)}
\def \[{\left[}
\def \]{\right]}

\def \Re {\text{Re}}

\usepackage[normalem]{ulem}

\begin{document}

\author{Gaoqing Cao}
\email{caogaoqing@mail.sysu.edu.cn}
\affiliation{School of Physics and Astronomy, Sun Yat-Sen University, Guangzhou 510275, China.}
\author{Shu Lin}
\email{linshu8@mail.sysu.edu.cn}
\affiliation{School of Physics and Astronomy, Sun Yat-Sen University, Guangzhou 510275, China.}

\title{Gravitational Wave from Phase Transition inside Neutron Stars}
\date{today}

\begin{abstract}
In this work, we propose a new source for gravitational wave (GW) radiation associated with the quantum chromodynamics (QCD) phase transition in the inner cores of neutron stars. The mechanism is based on the bubble dynamics during the first-order phase transition from nuclear matter to quark matter. We identify the characteristic frequency to be of order $\o_c\sim 10^6~{\rm rad/s}$ for this kind of sources and the strain magnitude ($h\sim 10^{-24}$ for a neutron star at a distance of $0.1~{\rm Mpc}$) reachable by future GW detectors. The GW spectra are shown to be useful to check the transition nature at high baryon chemical potential as well as to constrain the radius and density of the inner cores, which are still indistinct up to now.
\end{abstract}
\maketitle 

\sect{Introduction}%
The composition and structure of neutron stars (NSs) is one of the most intriguing but difficult questions in modern astronomy. Over the past decades, significant efforts have been devoted to establishing possible existences of quark matter in the inner cores of NSs \cite{Ivanenko:1969gs,Itoh:1970uw,Baym:1976yu,Benic:2014jia,Sotani:2001bb,Marranghello:2002yx}. However, the nature of inner cores remains a big puzzle owing to the difficulties in accessing that region either observationally or theoretically: On one hand, the outer cores of NSs are so dense and thick that the electromagnetic signals are unable to escape; on the other hand, the first principle lattice QCD calculation of the equation of state remains hard at high baryon density due to the sign problem~\cite{DElia:2012zw,Bazavov:2017dus}. The successful detections of GW from both binary black holes mergers~\cite{Abbott:2016blz,Abbott:2016nmj,Abbott:2017vtc} and binary neutron stars merger~\cite{TheLIGOScientific:2017qsa} by LIGO and Virgo open a new window into the interior of neutron stars. It turns out that the clean and precise GW signals offer great constraints on the ambiguous equation of state of NSs \cite{Hotokezaka:2011dh,Takami:2014zpa,Annala:2017llu,Annala:2017tqz,Li:2018ayl,Zhu:2018ona,Zhang:2018bwq,Most:2018eaw}. In this work, we propose a new source for GW, which is associated with the QCD phase transition in the inner cores of NSs.

As a matter of fact, the underlying mechanism of GW generation from the new source is analogous to the one in early universe, where GW radiation is induced by first-order phase transitions (FPTs) \cite{Turner:1990rc,Kosowsky:1992rz,Kosowsky:1991ua,Kosowsky:1992vn,GarciaBellido:2007af,Huber:2008hg,Hindmarsh:2013xza,Kalaydzhyan:2014wca,Chen:2017cyc}. However, the FPT here is driven by density rather than by temperature as in early universe: as a neutron star undergoes quick gravitational collapse during supernova explosion, the baryon chemical potential $\m_B$ of inner core eventually exceeds the critical value and then turns the nuclear matter into quark matter. The transition is usually thought to restore chiral symmetry and found to be of first-order in chiral effective models~\cite{Zhuang:1994dw,Schaefer:2006ds,Fukushima:2013rx}. Above the critical $\mu_B$, the FPT proceeds via the nucleation and expansion of quark matter bubbles inside the metastable nuclear matter. Meanwhile, a huge amount of latent heat is released, part of which is finally converted to GW radiation through bubble collisions~\cite{Kosowsky:1992rz,Kosowsky:1991ua,Kosowsky:1992vn}. Note that this kind of GW is transient and differs much from the periodic quasi-normal mode generated by hybrid NSs~\cite{Sotani:2001bb,Marranghello:2002yx,Fu:2017mcw}. As the GW carries specific information of the underlying FPT, it may shed light on the properties of inner cores once detected in the future. Throughout the work, the following conventions are used $\hbar=c=1$.

\sect{First-order phase transition}%
In order to study phase transition in the cores of neutron stars, we simply adopt the renormalizable quark-meson model~\cite{Schaefer:2006ds}. The Lagrangian density is given by
\begin{align}
{\cal L}_{QM}=&{1\over2}\Big[(\partial_\mu\sigma)^2+(\partial_\mu\boldsymbol{\pi})^2\Big]-{\lambda\over4}\Big(\sigma^2+\boldsymbol{\pi}^2-\upsilon^2\Big)^2+c~\sigma\nonumber\\
&+\bar{q}\Big[i\slashed{\partial}+{\mu_B\over N_c}\gamma^0-g\Big(\sigma+i\gamma^5\boldsymbol{\tau\cdot\pi}\Big)\Big]q,
\end{align}
where $q(x)=(u(x),d(x))^T$ denotes the two-flavor quark field with color degrees of freedom $N_c=3$, $\mu_B$ is the baryon chemical potential and ${\boldsymbol\tau}=(\tau^1,\tau^2,\tau^3)$ are Pauli matrices in flavor space. The linear term $c~\sigma$ breaks chiral symmetry explicitly and we can verify that the Lagrangian has exact chiral symmetry in the chiral limit $c=0$. The model parameters of the mesonic sector $\l,\upsilon,c$ are fixed by the sigma mass $m_\s=660~{\rm MeV}$, pion mass $m_\pi=138~{\rm MeV}$ and decay constant $f_\pi=93~{\rm MeV}$, and the quark-meson coupling constant is determined by $m_q^v\equiv gf_\pi=m_\s/2$ in the chiral symmetry breaking phase (for the stability of nucleons, $N_cm_q^v>m_N$)~\cite{Schaefer:2006ds}.

\begin{figure}
  \centering
  \includegraphics[width=7cm]{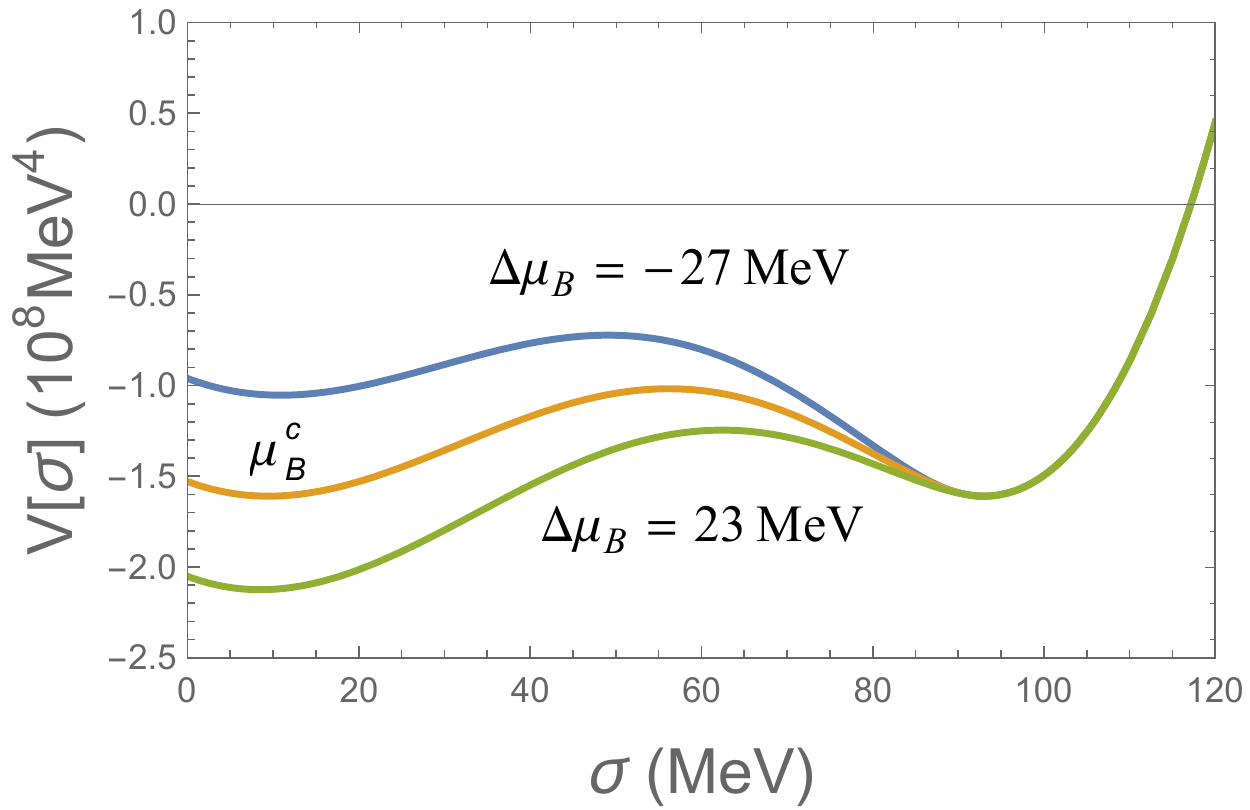}
  \caption{
(color online) The effective potential $V[\s]$ as a function of constant $\s$ at three specific values of baryon chemical potential: critical $\mu_B^c$ (middle), subcritical $\Delta\m_B=-27~{\rm MeV}$ (upper) and supercritical $\Delta\m_B=23~{\rm MeV}$ (lower), where $\Delta\m_B=\mu_B-\mu_B^c$.}
  \label{fig_Veff}
\end{figure}

For the study of cold neutron stars, we stick to the zero temperature and finite baryon chemical potential case. In reality, the isospin density is also large inside NSs but will be neglected here for simplicity. As the first step for illustrative purpose, we neglect the pion contributions to the bubble dynamics and assume the derivative terms $\partial_{\mu_1}\cdots\partial_{\mu_n}\sigma(x)$ have already been renormalized. More rigorous exploration can follow the pioneering works of Friedberg-Lee soliton model (though for hadrons)~\cite{Friedberg:1976eg,Friedberg:1977xf}, where the coupled equations for bosons and fermions should be solved consistently. Then, by integrating out the quark field and dropping the vacuum term as in the standard procedure~\cite{Schaefer:2006ds}, the Lagrangian density can be bosonized as a functional of $\s(x)$, that is, ${\cal L}_\s=\frac{1}{2}\(\pd_\mu\s\)^2-V[\s]$ with the effective potential
\begin{align}\label{Leff}
V[\s]=&\frac{\l}{4}\(\s^2-v^2\)^2-c\s-\frac{1}{12\pi^2}\Big[2p_F^3\m_B-3\Big(p_F\m_Bm_q^2\no
&-N_cm_q^4\cosh^{-1}\frac{\m_B}{N_cm_q}\Big)\Big]\th\left({\mu_B\over N_c}-m_q\right).
\end{align}
Here, we have defined the quark mass $m_q\equiv g\s(x)$ and the Fermi momentum $p_F\equiv\sqrt{(\m_B/N_c)^2-m_q^2}$. In the constant $\s$ scenario, the effective potential $V[\s]$ is depicted in Fig.~\ref{fig_Veff} for three specific values of $\m_B$. As can be seen, there are usually two minima for $V[\s]$: one at $\s\gg0$ corresponding to the chiral symmetry breaking phase and the other at $\s\sim0$ corresponding to chiral symmetry restoration phase (quark matter)~\cite{Klevansky:1992qe}. When we increase $\m_B$ from the subcritical case (upper curve) to the supercritical one (lower one), the true vacuum (general minimum of $V[\s]$) jumps from $\s\gg0$ to $\s\sim0$. This is the distinct feature of symmetry related first-order phase transition and the critical value is found to be $\m_B^c=957~{\rm MeV}$ for the chosen parameters. 

Thus, $\m_B$ has to be supercritical in order that the chiral transition from nuclear to quark matter can happen. In reality, $\m_B$ is not a constant in NSs but rather has a profile from center out, because we've known that the baryon density changes with the radius~\cite{Li:2008gp}. We just choose a simple profile: $\m_B=\m_{B1}\th(R_c-r)+\m_{B2}\th(r-R_c)$ with $\m_{B1}>\mu_B^c$ and $\m_{B2}<\mu_B^c$. Only the inner cores of neutron stars with $r<R_c$ are relevant for bubble nucleation and the radius is usually found to be $R_c=0-3~{\rm km}$ from theoretical studies~\cite{Weber:2004kj}.

\begin{figure}
	\centering
	\includegraphics[width=13.5em]{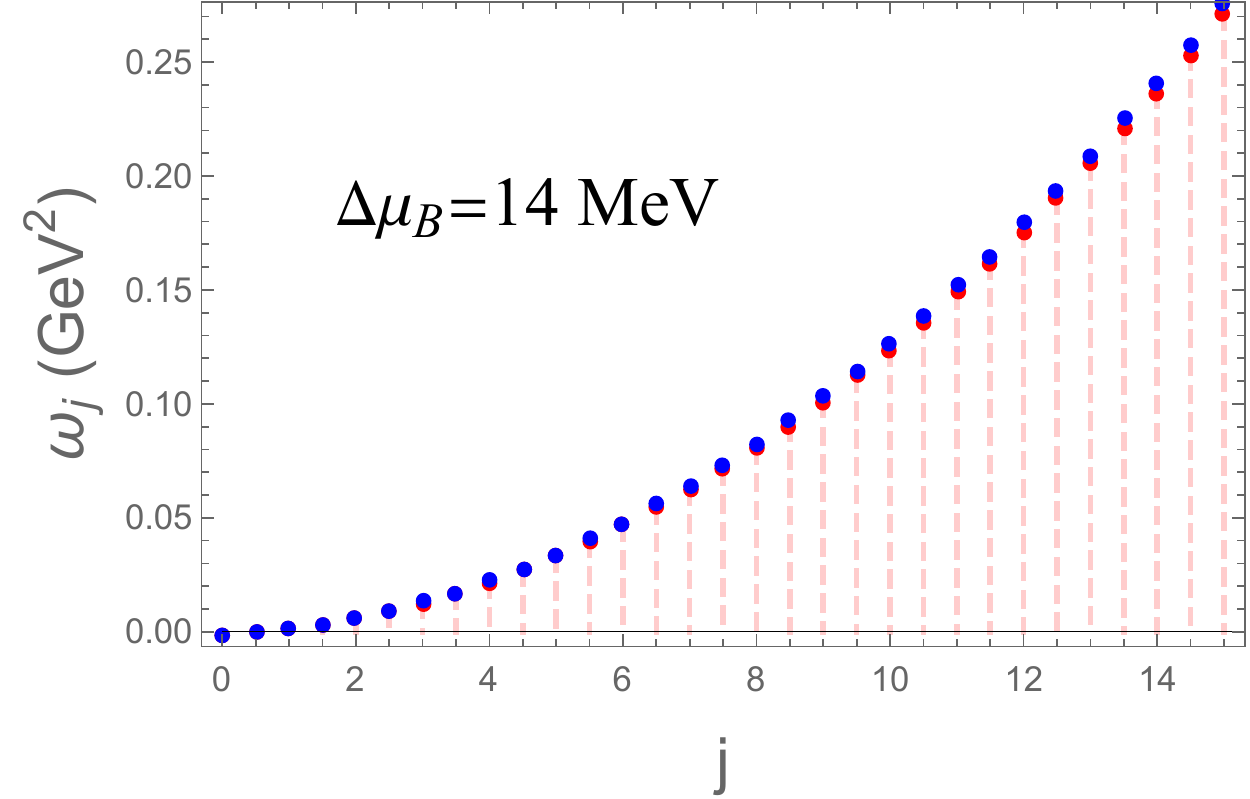}
	\includegraphics[width=12.8em]{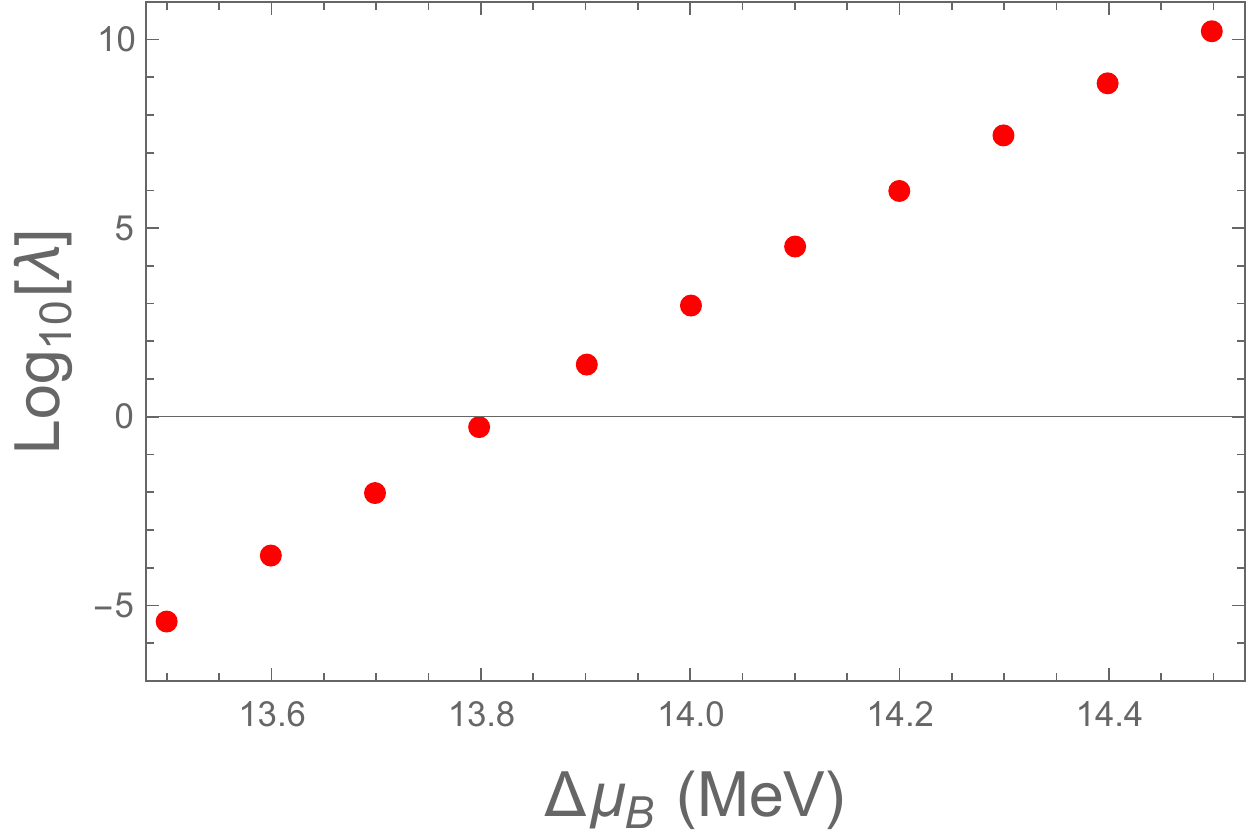}
	\caption{(color online) Left: Comparison of the exact numerical eigenenergy $\o_j$ of quantum fluctuations (blue dots) with those from thin wall approximation (red dots) at a chosen chemical potential $\D\m_B=14~{\rm MeV}$. Right: The average total nucleated bubbles $\lambda$ as a function of $\m_B$ for the core volume $V_c=\frac{4\pi}{3}R_c^3$ and time $T=R_c$. The radius $R_c$ is taken to be $1~{\rm km}$.}
	\label{fig_GVT}
\end{figure}

\sect{Bubble dynamics}
A significant application of FPT to neutron stars is the existence of the super-compressed phase in the cores -- gravity does the necessary work. Then, the phase transition from nuclear to quark matter will proceed through the nucleation and expansion of quark-matter bubbles inside the nuclear matter, which in the following will be referred to as true and false vacuum, respectively. The bubbles are nucleated through quantum tunneling effect and a single bubble is an $O(4)$ symmetric solution to the following equation of motion in Euclidean space-time~\cite{Coleman:1977py}:
\begin{align}
\frac{\pd^2\s}{\pd\r^2}+\frac{3}{\r}\frac{\pd\s}{\pd\r}=\frac{\pd V}{\pd\s},
\end{align}
where $\r=\sqrt{(t_E-t_{E0})^2+({\bf x}-{\bf x}_0)^2}$ is the Euclidean distance from the bubble center at $(t_{E0},{\bf x}_0)$. The bubble solution interpolates between the true vacuum in the center and the false vacuum outside, and bubbles are generated randomly across the whole region of false vacuum. 
Their number density is determined by the tunneling probability density $\G$, which can be evaluated from the Euclidean action and quantum fluctuations~\cite{Coleman:1977py,Callan:1977pt}. The explicit expression is $\G=A{B^2\over4\pi^2}e^{-B}$ with the coefficients
\begin{align}
 A=&\left|{{\rm Det}'(-\pd^2+V''[\s_b])\over {\rm Det}(-\pd^2+V''[\s_F])}\right|^{-1/2},\no
 B=&{2\pi^2\int_0^\infty \r^3d\r\left\{{1\over2}\({d\s_b\over d\r}\)^2+V[\s_b]-V[\s_F]\right\}},
\end{align}
where ${\rm Det}'$ denotes the determinant without zero modes, and $\s_b$ and $\s_F$ correspond to the bubble and false vacuum  solution, respectively. The exponent $B$ is positive definite because the bubble solution maximizes the Euclidean action~\cite{Coleman1985}. It is complicated to evaluate the prefactor $A$ in a renormalized way, which receives contribution from quantum fluctuations on top of $\s_b$~\cite{Coleman1985}. However, in the thin wall approximation (TWA)~\cite{Coleman:1977py,Stone:1975bd}, the prefactor is simply $A={4\over{\pi^2 R_b^4}}e^{\zeta_R'(-2)}$ with $\zeta_R(x)$ Riemann's zeta function~\cite{Garriga:1993fh} and the bubble radius $R_b$ can be determined by minimizing $V''[\s_b(\r)]$. On the left panel of Fig.\ref{fig_GVT}, we illuminate numerical results for dozens of lowest eigenenergy of the quantum fluctuations. They are in precise agreement with the analytic ones from the TWA: $\o_j=(4j^2+4j-3)R_b^{-2}$ with $j=0,1/2,1,\dots$ and thus verify the validity of TWA for our study.

After nucleation, bubbles start to expand and eventually collide with each other. These processes simply follow the classical equation of motion and the whole false vacuum will be occupied by the bubbles in the end. The speed of bubble walls was found to approach the light velocity $c$ quickly, with the wall width shrinks quickly~\cite{Kosowsky:1991ua}. This again justifies the validity of TWA for GW generation in the following and the time scale of FPT can be roughly evaluated to be $T\sim R_c$.

\sect{Gravitational wave radiation}%
As the nucleated bubbles expand and collide, the variation of stress tensor induces GW radiation from the core. The stress tensor is related to the bubble configuration $\sigma(x)$ as~\cite{Kosowsky:1992rz,Kosowsky:1991ua}.
\begin{align}
T_{ij}({\hat {\bf k}},\o)=\frac{1}{2\pi}\int_0^\infty dt e^{i\o t}\int d^3x\pd_i\s\pd_j\s e^{-i\o{\hat {\bf k}}\cdot{\bf x}},
\end{align}
where ${\hat {\bf k}}$ is the unit wave vector pointing from the neutron star to detector and $\o$ is the angular frequency of GW. The stress tensor will be evaluated by adopting the envelope approximation, which was shown to be in good agreement with the exact numerical evaluation~\cite{Kosowsky:1992vn}. The approximation is based on two simplifications: (i) the bubbles expand spherically with speed of light and do not interfere with each other; (ii) the overlapped regions of bubbles and the parts exceeding the boundary of false vacuum are excluded in the integration. As a result, the stress tensor is simply given as
\begin{align}\label{Tij}
T_{ij}({\hat {\bf k}},\o)=&\frac{\varepsilon_{\rm v}}{6\pi}\int_0^\infty dt~e^{i\o t}\left[\sum_{n=1}^{N}(t-t_n)^3e^{-i\o{\hat {\bf k}}\cdot{\bf x}_n}\right.\no
&\left.\times\int_{S_n}d\O~e^{-i\o{\hat {\bf k}}\cdot{\bf x}}{\hat{\bf x}}_i{\hat{\bf x}}_j\right],
\end{align}
where $n$ enumerates number of bubbles, $\O$ is the solid angle of each bubble wall, and $t_n$ and ${\bf x}_n$ are nucleation time and center location of the $n$-th bubble, respectively. The overall magnitude of $T_{ij}$ is set by the latent heat density $\varepsilon_{\rm v}$, the energy density difference between the true and false vacuum.

The resultant GW strain is then given by
\begin{align}\label{hij}
h_{ij}(t)=\frac{8G}{L}\Re\int_0^\infty\!\!\!\! d\o~ e^{-i\o (t\!-\!L)}\!\left[T_{ij}\!-\frac{g_{ij}}{2}T^\mu_\mu\right]({\hat {\bf k}},\o),
\end{align}
where $L$ is the distance from the neutron star to detector. To be specific, we choose ${\bf \hat k}$ along $z$-axis. Then, the stress tensor components relevant for observation are only $T_+\equiv {1\over2}(T_{xx}-T_{yy})$ and $T_\times\equiv T_{xy}$ with $+$ and $\times$ denoting different polarization modes. Another important observable is the differential GW energy spectrum ${\cal E}_{GW}\equiv\frac{\partial^2E_{GW}}{\partial\o \partial\O_{ob}}$, where $E_{GW}$ is the GW energy and $\O_{ob}$ is the observational solid angle~\cite{Weinberg1972}. In our convention, it can be simply split into two parts: ${\cal E}_{GW}={\cal E}_{GW}^++{\cal E}_{GW}^\times$, where ${\cal E}_{GW}^{+/\times}$ is defined as
\begin{align}\label{E_spec}
{\cal E}_{GW}^{+/\times}=\frac{4G\o^2}{\pi}\left|T_{+/\times}({\hat {\bf k}},\o)\right|^2.
\end{align}

Before we proceed to the realistic evaluations of $h_{+/\times}$ and ${\cal E}_{GW}^{+/\times}$, we need to determine the space-time coordinates of all bubbles nucleated during the FPT first. For the chosen size of the super-compressed region, the average bubble number $\lambda\equiv\G V_cT$ is found to be very sensitive to $\m_B$, see the right panel of Fig.\ref{fig_GVT}. We concentrate on the case $\Delta\mu_B\sim 13.8~{\rm MeV}$ below which nucleation is highly suppressed, and both few- and many-bubble scenarios will be considered in the following. Assuming individual bubble to be nucleated independently, the total number of bubbles $k$ follows the Poisson distribution rule $P(k)=e^{-\l}\frac{\l^k}{k!}$~\cite{Guth:1980zm}. Physically, nucleation of bubbles only occurs inside the false vacuum; nevertheless, we can still allow nucleation inside the true vacuum, i.e. other bubbles nucleated before. The point is that the later has no effect on the stress tensor under the envelope approximation, but the implementation is much easier than the equivalent multi-time-step evolution one~\cite{Kosowsky:1992vn}. With $k$ randomly generated from the Poisson distribution, the space-time coordinates of all the bubbles are also randomly generated in the space volume $V_c$ within time cutoff $T$.

\begin{figure}
	\centering
	\includegraphics[width=13em]{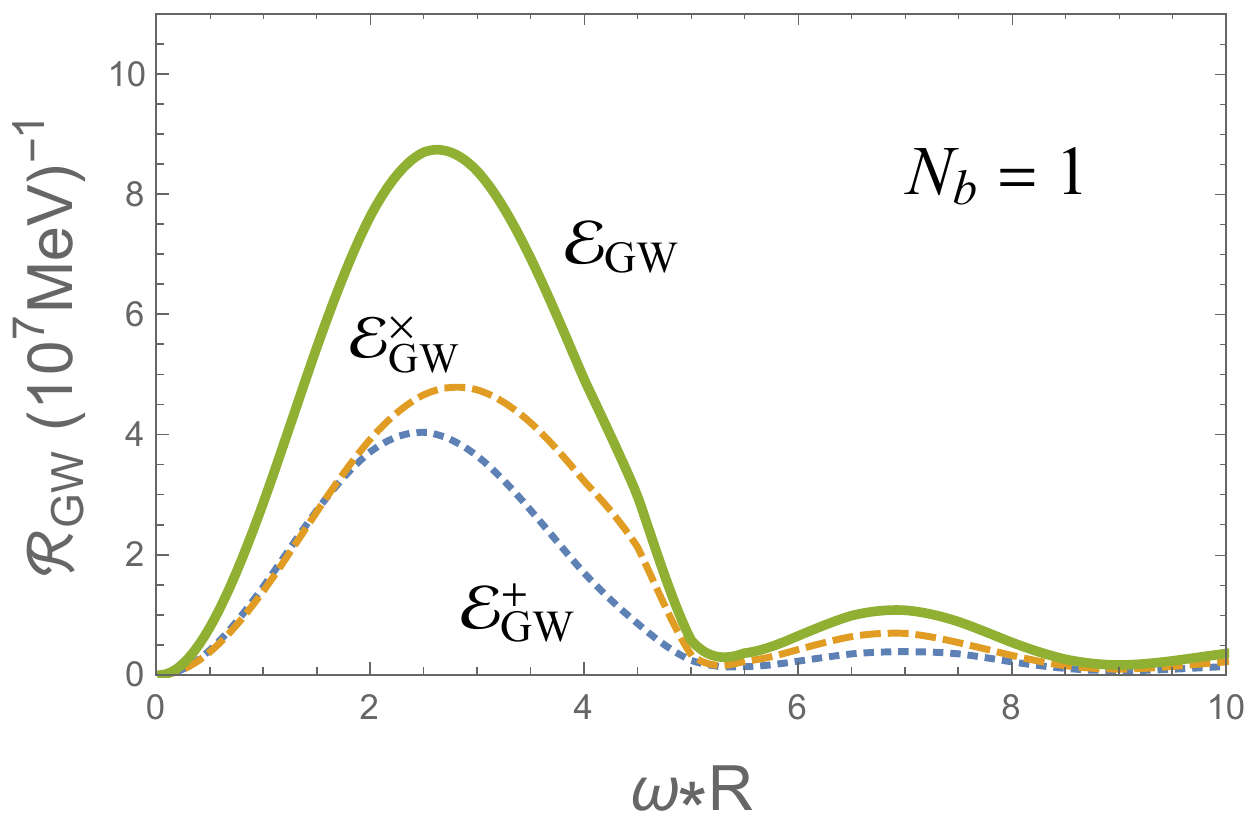}
	\includegraphics[width=12.5em]{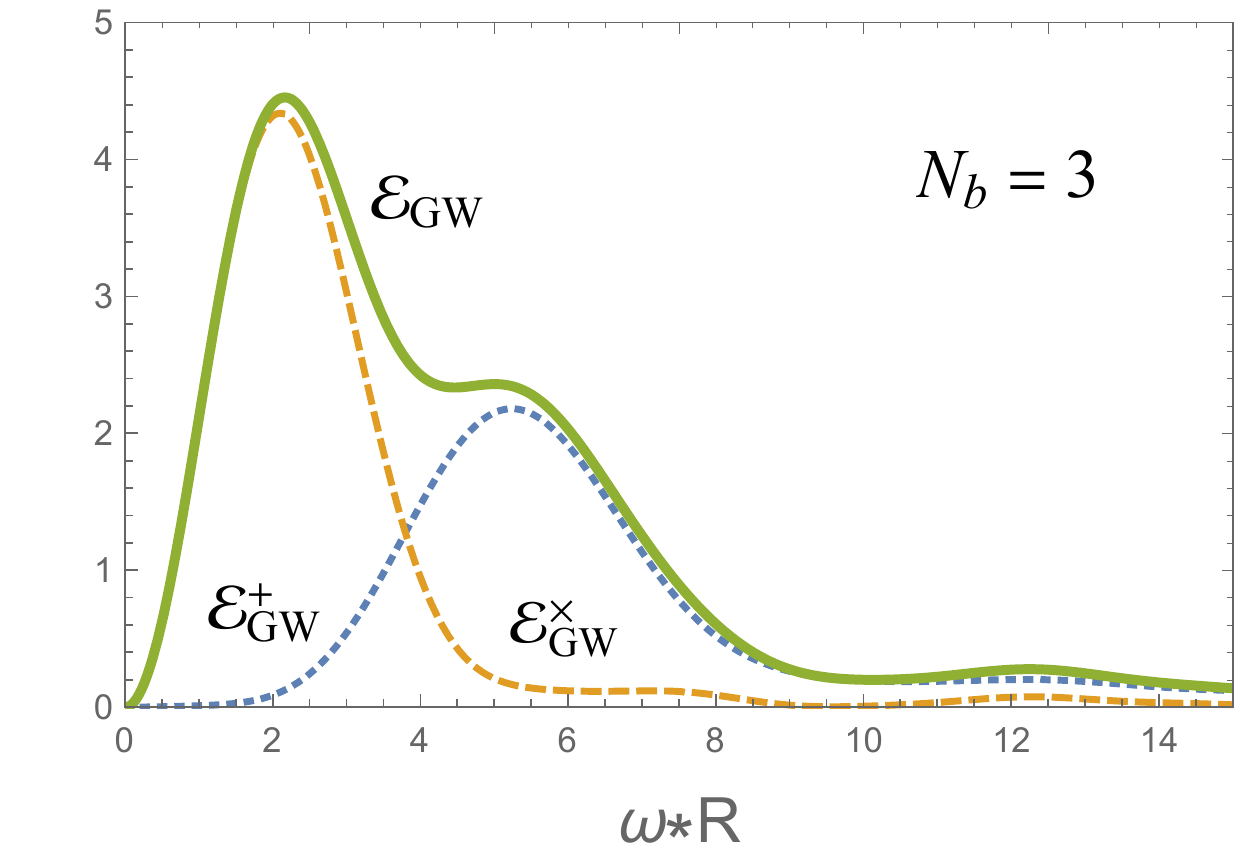}
	\includegraphics[width=13em]{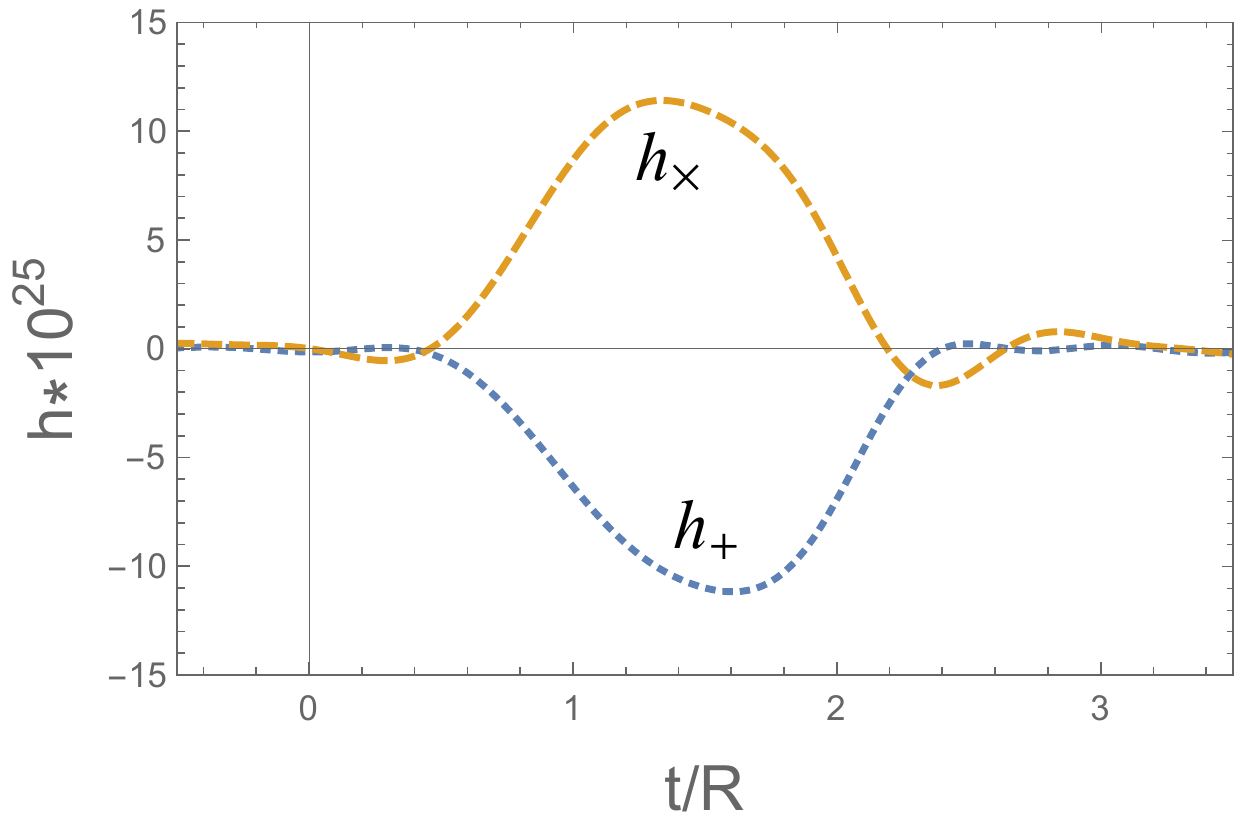}$\!\!$
	\includegraphics[width=12.8em]{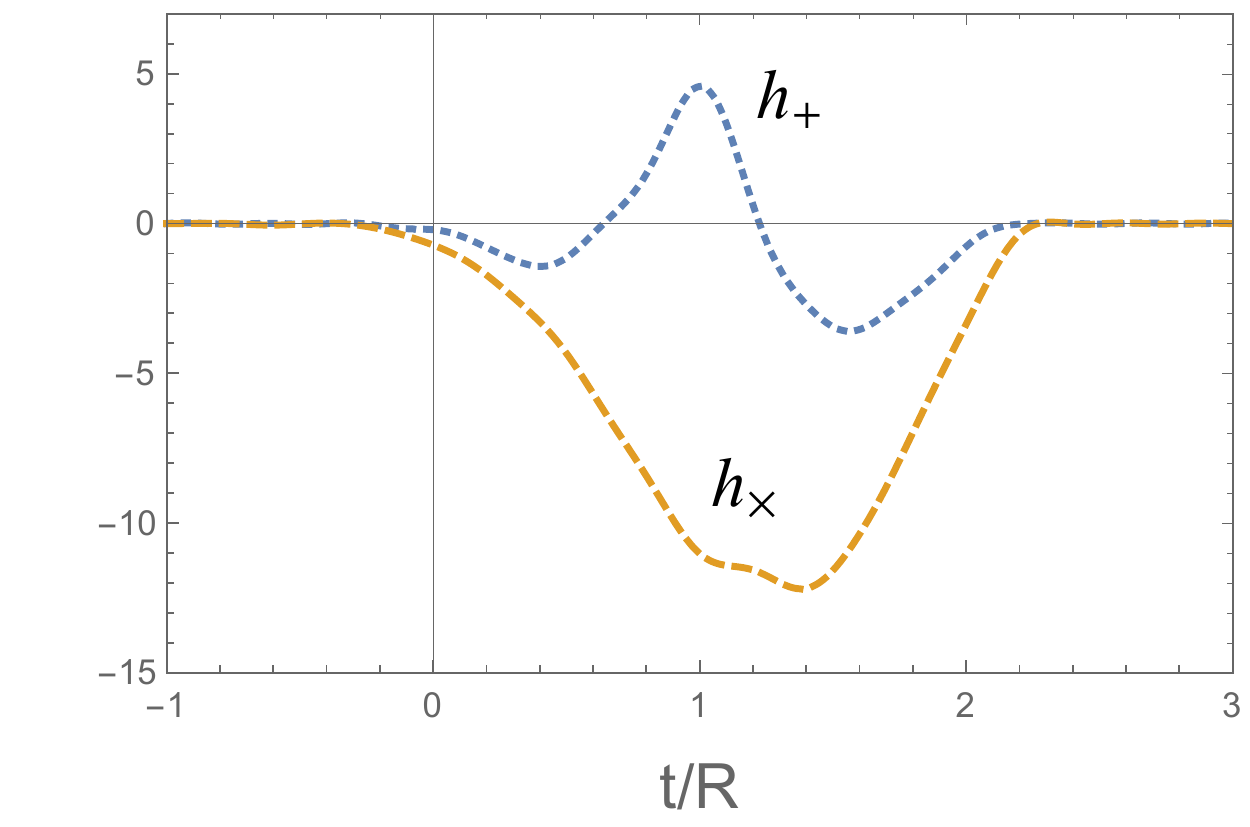}
	\caption{(color online) The GW energy spectra ${\cal R}_{GW}={\cal E}_{GW}/E_v$ (upper panels) and strains $h$ (lower panels) for the few-bubble scenario with bubble number $N_b=1$ (left) and $N_b=3$ (right). Here, the blue dotted and orange dashed lines correspond to the $+$ and $\times$ polarization modes, respectively.}\label{fig_one}
\end{figure}

We start with the one-bubble evolution since this is the simplest and most representative case for few-bubble scenario. For FPTs in early universe, at least two bubbles are required to generate GW radiation~\cite{Kosowsky:1992rz,Kosowsky:1991ua}. However, in our case, one bubble can also have such an effect because the boundary of false vacuum provides another big bubble (though not expanding). The results for a typical one-bubble configuration are illuminated on the left panel of Fig.\ref{fig_one} with the total latent energy $E_v=\varepsilon_{\rm v}V_c$. A significant feature is that there is only one obvious extreme for either the GW energy spectrum or strain of each polarization mode, and the strains are nearly in phase with each other except for a possible sign difference. The sign difference can be understood as follows: Since GW polarization modes have helicity $2$, rotation of the bubble center around $z$-axis by an angle $\varphi$ rotates $(h_+,h_\times)$ like a vector but by an angle $2\varphi$. Hence, rotation would not change the relative phase between the two modes, but can change their signs. The characteristic frequency (CF) for the energy peak can be estimated from the exponent in Eq.(\ref{Tij}) near the end of the FPT, when the factor $(t-t_1)^3$ is maximized. For a bubble nucleated at spherical coordinate $(r_1,\theta_1,\varphi_1)$, we find the transition time $T\simeq R+r_1$ and the end point ${\bf x}\simeq (T,\pi-\theta_1,\pi+\varphi_1)$. Then, the CF should be inversely proportional to the effective time $T_1\equiv T-{\hat{\bf k}}\cdot{\bf x}_1-{\hat{\bf k}}\cdot{\bf x} \simeq T+R\cos\theta_1$. By fitting several one-bubble results, we find $\o_1\simeq 3.8/T_1$ works surprisingly well as the two bubble case~\cite{Kosowsky:1991ua}. For illumination, we also present the results for $3$ bubbles on the right panel of Fig.\ref{fig_one}. The CF can be roughly evaluated by $\min(\o_1,\o_2,\o_3)$ and the deviation from numerical calculations should be attributed to the interference among bubbles.

\begin{figure}
  \centering
  \includegraphics[width=13em]{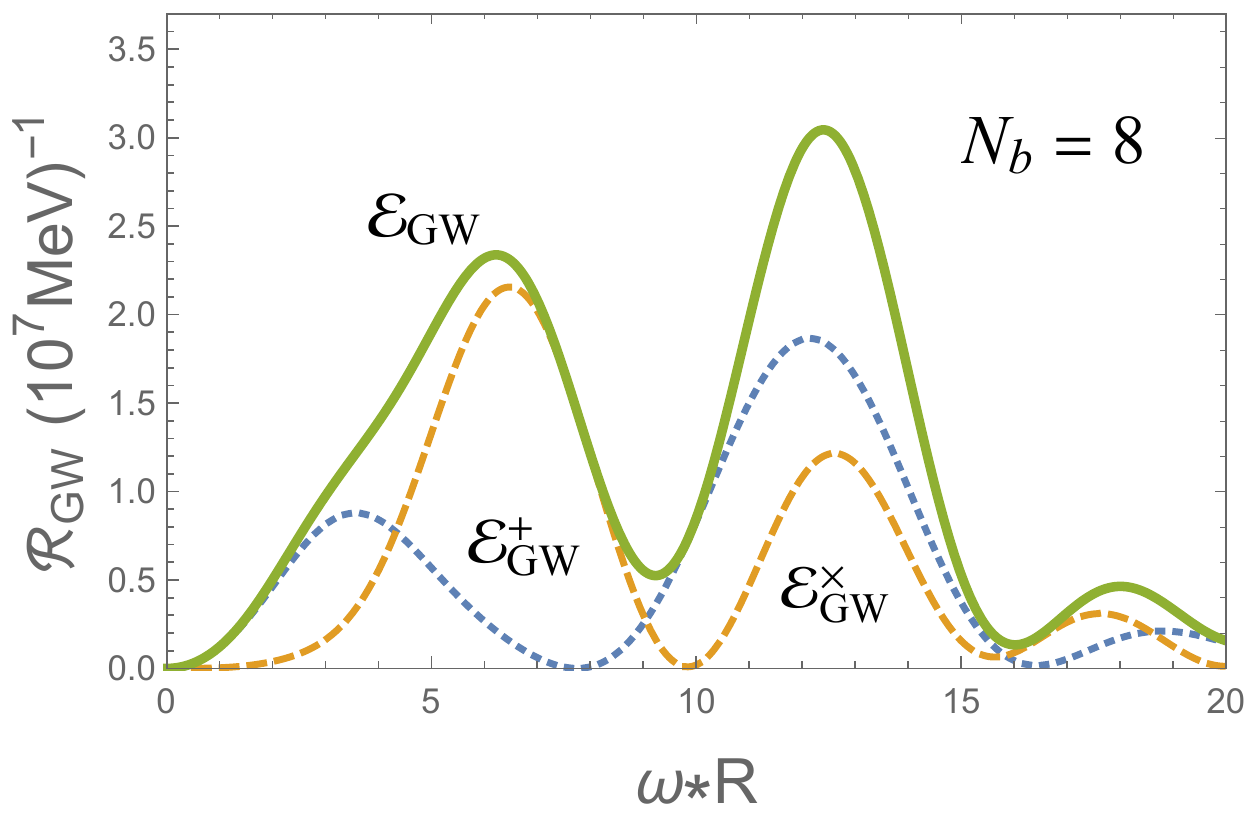}
  \includegraphics[width=13em]{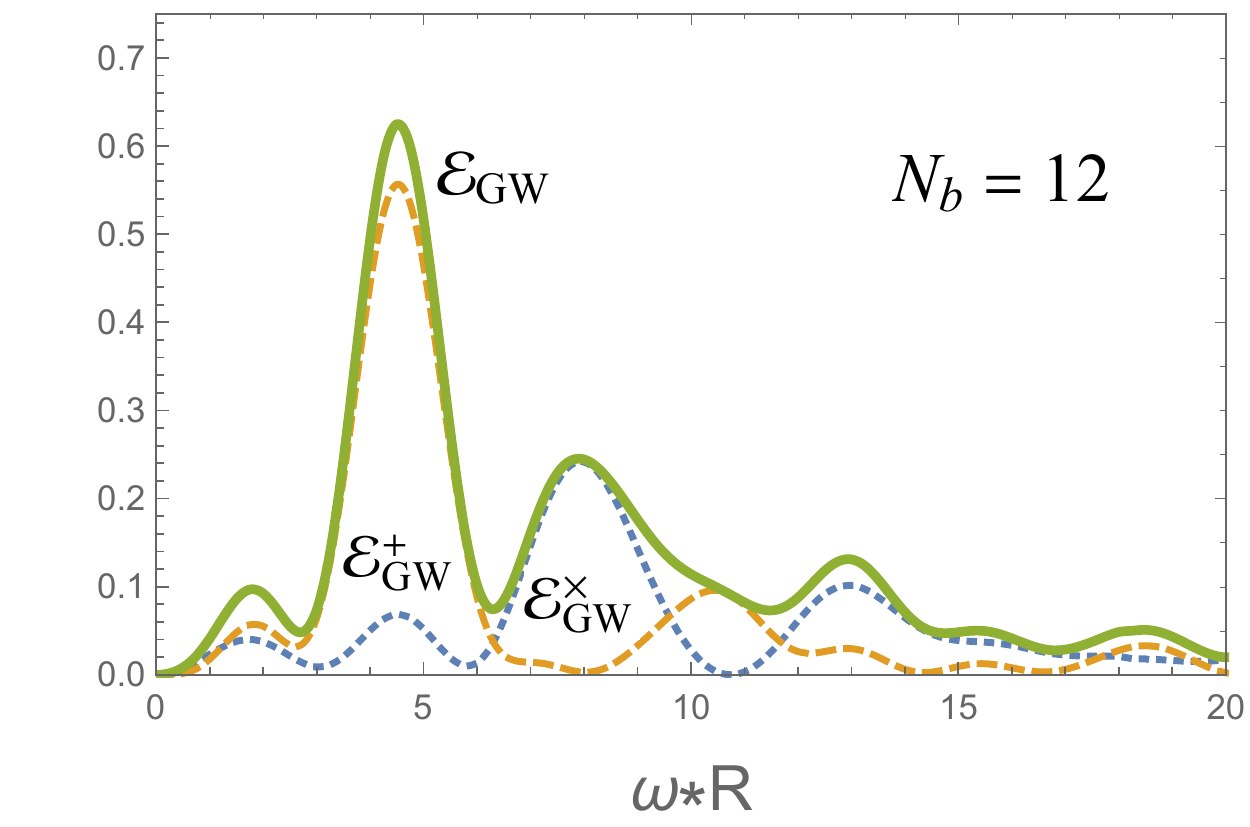}
  \includegraphics[width=13em]{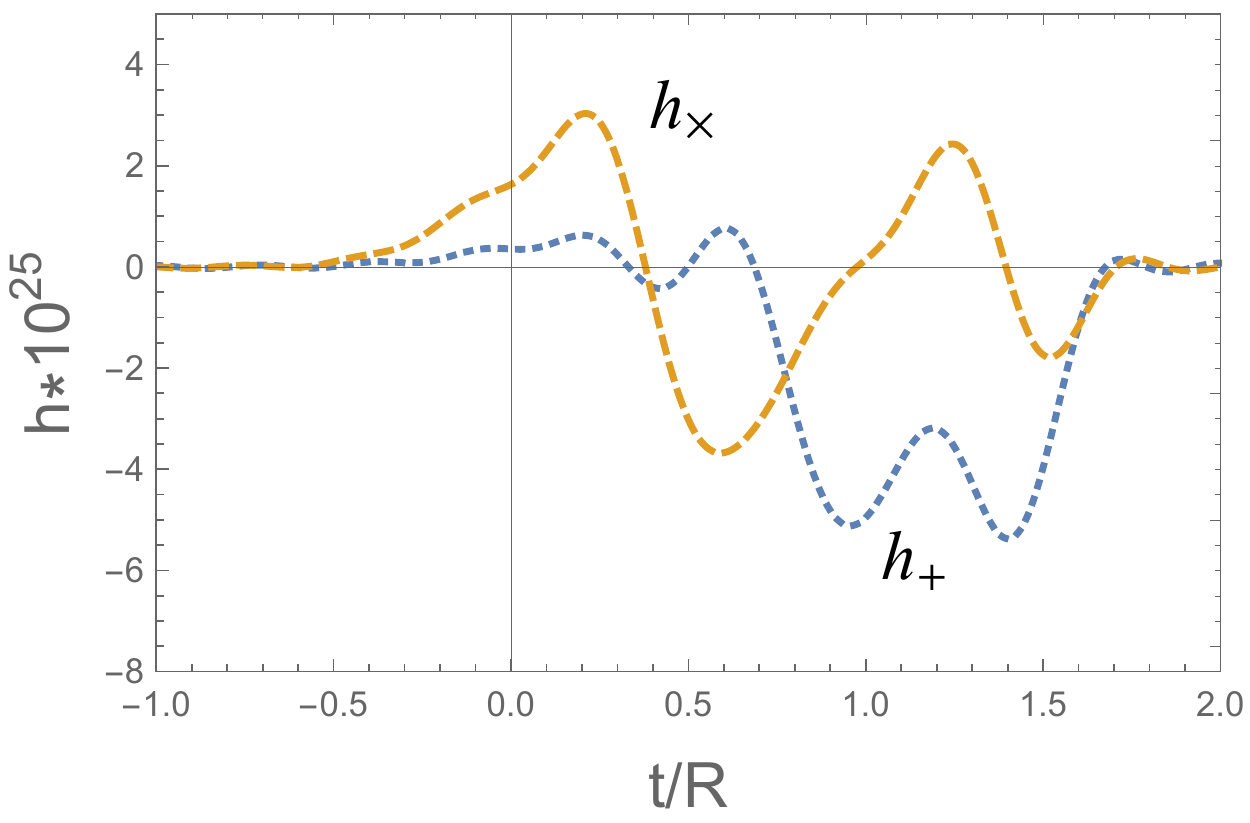}
  \includegraphics[width=13em]{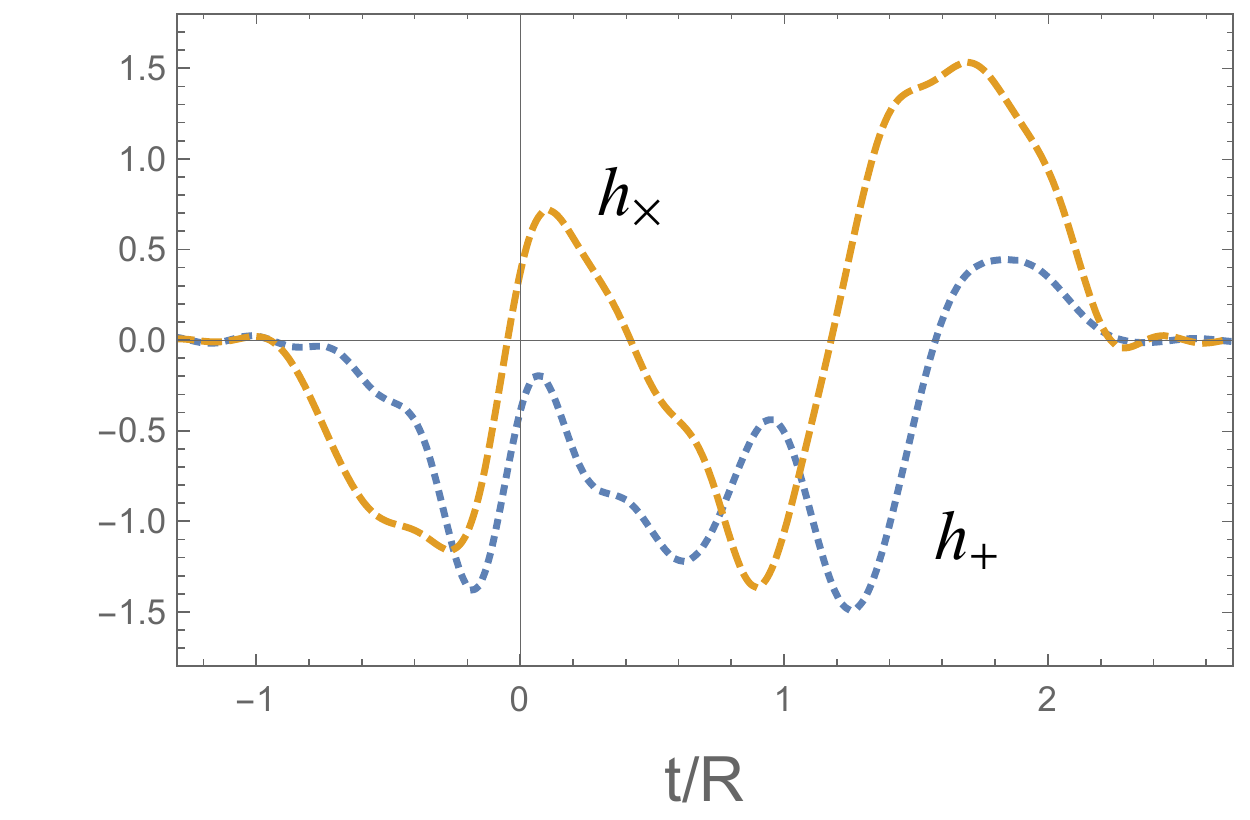}
  \caption{(color online) The GW energy spectra ${\cal R}_{GW}$ (upper panels) and strains $h$ (lower panels) for the many-bubble scenario with bubble number $N_b=8$ (left) and $N_b=12$ (right). The notations are the same as in Fig.\ref{fig_one}.}\label{fig_two}
\end{figure}

We carry out the same analysis for the scenario of many-bubble evolution. The corresponding GW energy spectra and strains are shown in Fig.\ref{fig_two} for typical configurations with $8$ and $12$ bubbles. For these cases, multiple obvious extremes can be found in both observables, the magnitudes of which are smaller than the counterparts in the few-bubble scenario. On the other hand, the energy spectra/strain spans a wider range in frequency/time respectively, see Fig.\ref{fig_one}.
In principle, more bubbles will involve more CFs $\omega_n$ for different bubbles, thus more obvious extremes are able to be produced in the radiation curves. It can be verified that the extremes of the two polarization modes do not necessarily coincide with each other now and the in-phase property from the one-bubble case is also lost.

\sect{Discussions}
The GW radiation obtained above can be a very useful probe for the nuclear/quark matter phase transition in the inner cores of NSs. First of all, the characteristic frequency is approximately set by $\o_c\sim 2\pi/R_c$, which distinguishes itself from other sources of GW. Once detected, it serves as a clear evidence that QCD transition is of first-order at large $\m_B$. Secondly, the duration of the observed GW pulse $T_{ob}$ will strongly constrain the size of super-compressed region: For one-bubble configuration, we have $R_c/2<T_{ob}<4R_c$ with the upper and lower limits from the bubble nucleated at perigee and apogee, respectively. Thirdly, the latent heat density $\varepsilon_v$ can be extracted from the magnitude of GW strain, as can be seen from Eq.(\ref{Tij}) and Eq.(\ref{hij}). The information can then be transformed to the baryon density or chemical potential by following reliable QCD models. Finally, the particular wave forms of the GW can serve as an indicator of the scenarios for bubble nucleation: In-phase wave forms for the two polarization modes with one extreme would prefer few-bubble scenario, while out-phase wave forms with multiple extremes would support many-bubble scenario.

It is more practical to have a little numerical discussions. For the chosen radius $R_c=1~{\rm km}$ as in Fig.\ref{fig_one} and Fig.\ref{fig_two}, the CF can be roughly evaluated to $\o_c\sim6\pi\times 10^5~{\rm rad/s}$, which distinguishes itself by $3$ orders larger than that discovered in the merger of binary neutron stars~\cite{TheLIGOScientific:2017qsa}. For the $\m_B$ range shown on the right panel of Fig.\ref{fig_GVT}, the latent heat density is almost a constant in the quark-meson model: $\varepsilon_v=3.74\times10^7~{\rm MeV}^4$. This gives rise to the GW strains of order $10^{-25}-10^{-24}$ and total GW radiation energy $(10^{-13}-10^{-11})M_\odot$ ($M_\odot$ is the solar mass) for a near source with distance $L=0.1~{\rm Mpc}$, see Fig.\ref{fig_one} and Fig.\ref{fig_two}. The magnitudes of GW strains are still $2-3$ orders smaller than the threshold of the next generation GW detector 'Cosmic Explorer'~\cite{Evans:2016mbw}. However, if the size of inner cores is as large as $3~{\rm km}$ and the source is located in the active region between solar system and the center of Milky Way with $L<0.008~{\rm Mpc}$, the magnitude will increase by more than $3$ orders to be well reachable by several advanced detectors~\cite{Evans:2016mbw}. This is promising because the typical event rate for type II supernovae in spiral galaxies is about one event per $50-100$ years~\cite{Camenzind2007}. For the hypothetical quark stars with much larger radii $\sim 10~{\rm km}$~\cite{Ozel:2015fia}, the observational effect will be even more significant. 

In the end, we address the question of GW damping, mainly by the out cores composed of nuclear matter. The upper bound of GW absorption rate is given by~\cite{Baym:2017xvh}
\begin{align}
\gamma\lesssim 8\pi G \frac{P R_{\rm NM}}{\o},
\end{align}
where $P$ is the energy density of the nuclear matter and $R_{\rm NM}$ is the depth of the outer cores. The physical parameters can be reasonably evaluated as $P\sim 1~{\rm GeV/fm^3}$ and $R_{\rm NM}\sim6~{\rm km}$~\cite{Weber:2004kj}. Then, the absorption rate is constrained to $\gamma\lesssim0.03$ for $\o_c\sim6\pi\times 10^5~{\rm rad/s}$, which convinces us that the GW can escape the NSs to be detected by us.

\sect{Acknowledgments}
We are grateful for Yungui Gong and Rongfeng Shen for useful discussions. S.L. is supported by One Thousand Talent Program for Young Scholars and NSFC under Grant Nos 11675274 and 11735007.

\end{document}